\documentclass[lettersize,journal]{IEEEtran}
\usepackage{amsmath,amsfonts}
\usepackage{algorithmic}
\usepackage{array}
\usepackage{textcomp}
\usepackage{stfloats}
\usepackage{url}
\usepackage{verbatim}
\usepackage{graphicx}
\usepackage{makecell,multirow,diagbox}
\usepackage{bm}
\usepackage{color}

\def\BibTeX{{\rm B\kern-.05em{\sc i\kern-.025em b}\kern-.08em
    T\kern-.1667em\lower.7ex\hbox{E}\kern-.125emX}}
\usepackage{balance}
\usepackage{subfigure}
\begin{document}
\title{How to Define the Propagation Environment Semantics and Its Application in Scatterer-Based Beam Prediction}
\author{{Yutong Sun, Jianhua Zhang, Li Yu, Zhen Zhang, Ping Zhang}\\
\thanks{This work is supported by the National Science Fund for Distinguished Young Scholars (No.61925102), the National Natural Science Foundation of China (No.92167202), the National Natural Science Foundation of China (No.62101069),
the National Key R\&D Program of China (No.2020YFB1805002), and BUPT-CMCC Joint Innovation Center.
\par Yutong Sun, Jianhua Zhang, Li Yu, Zhen Zhang, and Ping Zhang are with the State Key Lab of Networking and Switching Technology, Beijing University of Posts and Telecommunications, Beijing 100876, China and Ping Zhang is also with Peng Cheng Laboratory, Shenzhen 518066, China (e-mail: sun\_yutong@bupt.edu.cn; jhzhang@bupt.edu.cn; li.yu@bupt.edu.cn; zhenzhang@bupt.edu.cn; pzhang@bupt.edu.cn).}}
\maketitle
\begin{abstract}
In view of the propagation environment directly determining the channel fading, the application tasks can also be solved with the aid of the environment information. Inspired by task-oriented semantic communication and machine learning (ML) powered environment-channel mapping methods, this work aims to provide a new view of the environment from the semantic level, which defines the propagation environment semantics (PES) as a limited set of propagation environment semantic symbols (PESS) for diverse application tasks. The PESS is extracted oriented to the tasks with channel properties as a foundation. For method validation, the PES-aided beam prediction (PESaBP) is presented in non-line-of-sight (NLOS). The PESS of environment features and graphs are given for the semantic actions of channel quality evaluation and target scatterer detection of maximum power, which can obtain 0.92 and 0.9 precision, respectively, and save over 87\% of time cost.
\end{abstract}


\section{Introduction}
\IEEEPARstart{W}{ith} the ever-increasing diverse scenes and communication requirements, predictive 6G Network with environment sensing enhancement is becoming promising \cite{b1}\cite{b2}. Powered by advanced sensing techniques, environment reconstruction can be deployed, and the detailed environment information \cite{b3}\cite{b4} enables the applications with precision improvement. However, the demanded new technologies with increasingly-high data rates require online predictions for dynamic environments, especially in non-line-of-sight (NLOS) scenarios.

Powered by natural language processing (NLP) and computer vision (CV) techniques that have lots of potential in processing intelligent tasks, semantic communication \cite{b5}-\cite{b7} has drawn significant attention, which mainly relies on semantic-based information conversion between different content to achieve efficient, intelligent interaction. Semantic communication focuses on the content between the transmitter (Tx) and receiver (Rx), which considers the difference between the meaning of the transmitted messages and that of recovered ones for different semantic tasks.  

Similar to the semantic pipeline of semantic communication, the propagation environment and application task also need semantic (one-to-many) mapping that carries the meaning rather than the object (one-to-one) mapping without understandable information because online applications do not always perform with every environment changing. In \cite{b8}\cite{b9}, the cluster nuclei is proposed by directly mapping the physical environment to the channel. For representing the propagation environment from different perspectives, the environment features and graph representations are proposed in \cite{b10} and \cite{b12}, which can assist the efficient channel prediction in dynamic environments. Hence, we believe that defining the propagation environment semantics (PES) by considering the different environment representations, i.e., propagation environment semantic symbols (PESS), is crucial for highly efficient prediction by leveraging the environment information directly.
 
In this paper, PES is defined as a PESS set, where the set is limited because of considering the channel with limited properties as the basis. The PESS are deconstructed environment representations at a semantic level for prediction applications. Therefore, semantic mapping can be built between the PES and applications, and the specific task can be implemented by the related PES. For beam prediction implementation in NLOS scenarios, the environment features and graph representations are considered the PESS for the semantic actions: channel quality evaluation and target scatterer detection. Compared with the method that predicts the beam indices and requires the extra process of beam searching \cite{b4}, the proposed method can provide the scatterer with maximum power directly and further empower other advanced techniques.

\section{Problem Formulation}
The wireless channel is the essential intermediate bridge for the semantic mapping between the propagation environment and the application tasks. Therefore, the task-oriented PES can be defined by the semantic deconstruction of the environment, which considers the channel properties. Then, the tasks can be employed based on the machine learning (ML) method, as shown in Fig. 1.
\begin{figure*}[!htb]
\centering
\includegraphics[width=0.94\textwidth]{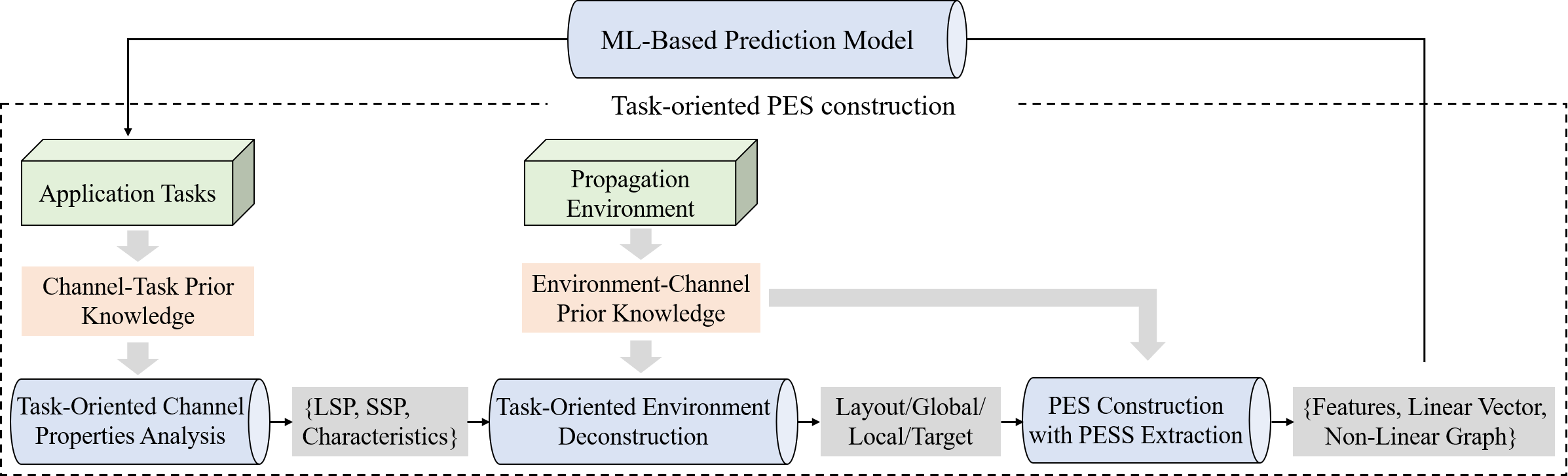}
\caption{Task-oriented PES construction process.}
\end{figure*}
\subsection{Task-Oriented Environment Deconstruction}
According to the channel-task prior knowledge, the concerned channel properties are impacted by different environmental information. Diverse properties or property groups are required to meet different application tasks, where the properties include large-scale parameters (LSP): path loss, delay spread (DS), azimuth angle spread of arrival (ASA), azimuth angle spread of departure (ASD), zenith angle spread of arrival (ZSA), zenith angle spread of departure (ZSD), small-scale parameters (SSP): power, delay, azimuth angle of arrival (AOA), azimuth angle of departure (AOD), zenith angle of arrival (ZOA), zenith angle of departure (ZOD), and characteristic: line-of-sight (LOS) blockage.

The environment is deconstructed to meet the applications directly at the semantic level. Following the environment-channel prior knowledge, the environment can be represented at large-scale and small-scale levels to meet the environment information requirement of different channel properties. The large-scale level includes the layout and global environment representations for LSP, and the small-scale level consists of local and target representations for SSP and LOS blockage.

\subsection{PES Construction with PESS Extraction}
The incurred semantics-based action of the application task depends on diverse background information, which decides how to interpret the intermediate information. As for the target task, environment information is not equally important to the specific semantics needs, so the task-related information should be abstracted instead of retaining all of it. As a result, the PES can be defined as the semantic variable that reflects the semantic changes. 

Unlike the physical environment without specific interactions between objects, the propagation environment should be described with the preset Tx, Rx, and propagation mechanism. The radio waves encountered with the scatterers can produce diverse propagation paths caused by various significant propagation mechanisms, such as LOS transmission, reflection, and diffraction. Therefore, the geometry relationship-correlated propagation mechanisms can be regarded as the considerable environment-channel prior knowledge for PESS extraction.

According to the propagation mechanisms, the essential geometric attributes that affect the paths include position, dimensions, and layout. Thereby, the fundamental PESS can be extracted as features $PESS_{feature}$, linear vectors $PESS_{vector}$, and non-linear graphs $PESS_{graph}$ to represent the environment characteristics or global structure at small-scale or large-scale level, as shown in Fig. 2. Its expandable when extra data form are raised and the basis PES can be presented as the set of the PESS, i.e.,
\begin{equation} 
PESS = \{PESS_{feature}, PESS_{vector}, PESS_{graph}\}.
\end{equation}

\section{PES for Beam Prediction}
A beam prediction case is given for method verification to show how the PES works on the prediction tasks.
\subsection{Task-oriented Channel Properties}
Beam prediction generally aims to improve the communication quantity by switching to a better target. Thus, two task actions should be considered for PES-based beam prediction: channel quality evaluation and target scatterer prediction, as shown in Fig. 2. Therefore, the respective channel properties can be analyzed according to the requirements.
\begin{figure*}[!htb]
\centering
\includegraphics[width=0.84\textwidth]{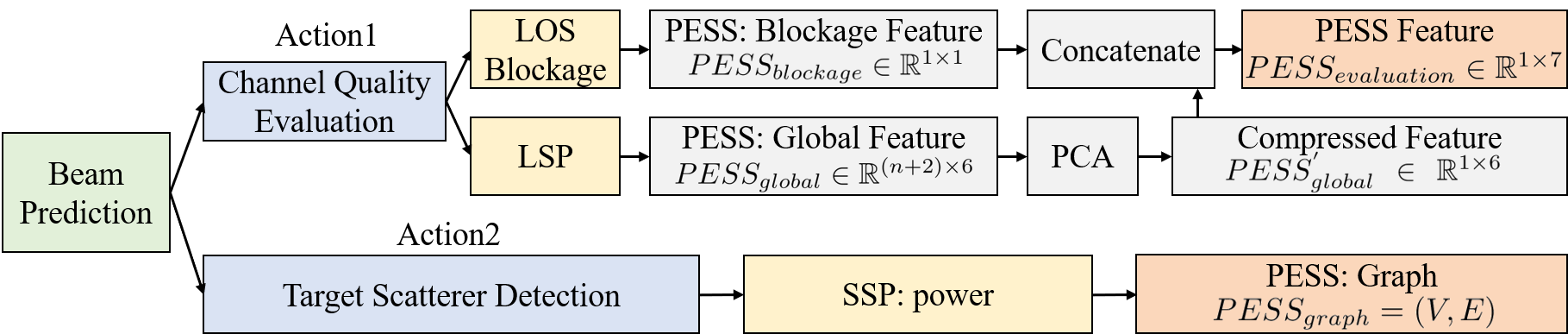}
\caption{The process of action decomposition and corresponding PESS extraction.}
\end{figure*}

The LSP can reflect channel quality evaluation related to the performance \cite{b10}, where the parameter types should not be considered in detail for coarse semantic mapping. In addition to the LSP, the blockage characteristic of LOS also has a significant impact, where the blockage would attenuate the performance. Therefore, the critical channel properties of quality evaluation action $A_{evaluation}$ can be expressed as
\begin{equation} 
A_{evaluation} = \{LSP, Blockage\}.
\end{equation}

Once it is determined that the current channel is unqualified, the target scatterer detection should be employed according to the maximum power, which is an SSP-related prediction problem. Hence, the channel properties of target scatterer prediction $A_{detection}$ can be denoted by
\begin{equation} 
A_{detection} = \{P_j, j \in J\},
\end{equation}
where $P_j$ is the power of the $j$-th path and total $J$ paths.
\subsection{Task-Oriented PES Construction}
According to the channel properties, the environment information influencing the concerned characteristics and parameters should be represented from global and structural aspects for two different semantic actions implementation. However, for practice applications, environment changes might cause propagation path changes but not significant statistics changes. Hence, the environment is represented for all the channel properties instead of each parameter or characteristic. The primary information of Tx, Rx, and internal scatterers are utilized as original data, which are environment-isolated information with no scene constraints.

{\bf{PESS of Features:}} For the simultaneous description of the blockage characteristics and statistical LSP, the blockage and global features are extracted and constructed into an exclusive PESS representation, as shown in Fig. 2.

The degree of LOS occlusion is leveraged as a PESS for blockage description. In practice, diverse methods can be utilized for PESS feature calculation according to the geometric relationship between the Tx, Rx, and the scatterer. There, we use the method mentioned in \cite{b10}, which obtains the blockage feature for each scatterer by describing the extent to which LOS and scatterers intersect. Specifically, the distance $d_i$ between the center point of $i$-th scatterer and the LOS is calculated according to the position coordinates. Then $b_i$ can be defined as the ratio of the $d_i$ and the width of the $i$-th scatterer $w_i$. Then the maximum blockage feature is selected as the environment-level PESS, which can be denoted as
\begin{equation} 
PESS_{blockage} = max \{b_1,b_2,\dots,b_n\},
\end{equation}
where there are $n$ internal scatterers.

As for the LSP representation, the PESS that contains the general environment information should be extracted. Because of the uncertain parameter requirement, the embedding feature should be utilized rather than the certain calculated feature. The matrix of original global representation $PESS_{global}$ is constructed by the information of the Tx, Rx, and internal scatterers. In view of the 3-dimensional coordinates vector of Tx and Rx while the 6-dimensional vector for scatterers, 0 paddings are utilized to fill the Tx and Rx row to deal with the inconsistency of dimensions. In which $PESS_{global} \in \mathbb{R}^{(n+2)\times6}$ for the environment sample with $n$ scatterers.

Therefore, the final PESS can be obtained by combining the matrix $PESS_{global}$ and the blockage feature value. To concatenate the two features of different dimensional, $PESS_{global}$ should be first converted into a 1-dimensional vector. The commonly used unsupervised dimensionality reduction algorithm: principal component analysis (PCA), is utilized for compression \cite{b11}. Hence, the compressed feature $PESS_{global}^{'} \in \mathbb{R}^{1\times6}$ is obtained. Finally, the blockage and global feature can be concatenated as a whole environment feature, i.e., $PESS_{evaluation} \in \mathbb{R}^{1\times7}$.

{\bf{PESS of Environment Graph:}} Unlike the overall channel properties, which the linear feature can represent, the environment layout needs a nonlinear representation. The graph structure data in non-Euclidean space is utilized to describe the structure information of the propagation environment. Specifically, the environment graph is constructed as the PESS for each scatterer that is to be classified. The graph is constructed by building edges of the pending scatterer node and other nodes to mark the pending scatterer \cite{b12}.

Therefore, let $PESS_{graph} = (V, E)$ denotes the graph with nodes $V$, edges $E$, and node feature vectors $X$. Where $V$ consist of Tx node $v_t$, Rx node $v_r$, and $n$ scatterer nodes for the graph of $n$ scatterers, i.e., $V=\{v_t, v_r, v_1, v_2, \dots, v_n\}$, as shown in Fig. 3. Meanwhile, $E$ can be expressed as $E = \{(v_t, v_p), (v_r, v_p),\dots,(v_n, v_p), n\neq p\}$. For Tx and Rx node, the position coordinates are used as feature vectors, that is, $X_t=(x_t, y_t, z_t)$ and $X_r=(x_r, y_r, z_r)$. The center coordinates $(x_i, y_i, z_i)$, long $l_i$, width $w_i$, and height $h_i$ formed the feature vector of $i$-th scatterer node, which can be expressed as $X_i = \{(x_i, y_i, z_i,l_i,w_i,h_i), i \in n\}$.
\begin{figure}[!htb]
\centering
\includegraphics[width=0.42\textwidth]{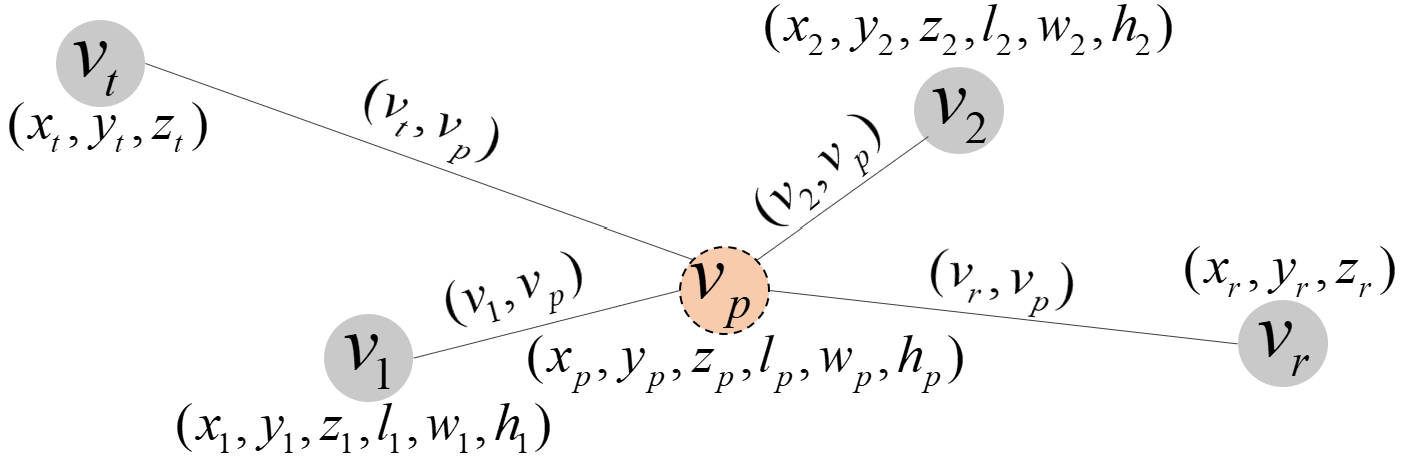}
\caption{The $PESS_{graph}$ of three scatterers and pending scatterer $v_p$.}
\end{figure}
\section{PESaBP}
After obtaining the essential PES of the beam prediction, the PESaBP can be implemented, i.e., channel quality evaluation and target scatterer detection can be achieved by leveraging the corresponding PES directly. The proposed PESS of environment features can predict the channel quality, and the constructed PESS of environment graphs can detect the target scatterer for beam prediction.
\subsection{PES-Based Channel Quality Evaluation}
As for dividing the quality into qualified and unqualified, the quality evaluation can be solved as a binary classification problem, differentiating the unqualified as class 0 and the qualified quality as class 1. In which the quality threshold can be set for diverse requirements according to the cumulative probability of the received power. According to the low-dimensional features for PES representation, the support vector machine (SVM) is built as the classification model rather than the neural networks requiring more learning cost with no significant accuracy gain.

The SVM \cite{b13} is a classic ML algorithm to maximize a particular mathematical function for a given collection of data that performs classification by constructing the hyperplane. The kernel function is the crucial calculation that enables the SVM to map the data from a low-dimensional space to a higher-dimensional space, which can be denoted by
\begin{equation} 
\left \langle a_1 \cdot a_2 \right\rangle \leftarrow K(a_i, a_j) = \left \langle \Phi(a_i) \cdot \Phi(a_j) \right\rangle,
\end{equation}
where $\Phi$ is a nonlinear function that maps the input space into the feature space and $K$ is the kernel function.
\par Four classical kernel functions are used for nonlinear model learning, including linear, polynomial, sigmoid, and radial basis kernels. In which the linear and polynomial kernel function can be described as
\begin{equation} 
K(a_i, a_j) = \left \langle a_i, a_j \right\rangle, K(a_i, a_j) = (1 + \left \langle a_1, a_2 \right\rangle)^d,
\end{equation}
where $d$ is the degree of the kernel function. The radial basis kernel can map the primitive features to infinite dimensions, which can be expressed as

\begin{equation} 
K(a_i, a_j) = exp(- \frac{\left \| a_i-a_j \right \|}{2 \sigma^2}).
\end{equation}
While the sigmoid kernel function comes from the neural network, which is generally denoted by
\begin{equation} 
K(a_i, a_j) = tanh(\gamma \left \langle a_1, a_2 \right\rangle +r),
\end{equation}
where the $\gamma$ and $r$ are the kernel parameters. 
\subsection{PES-Based Target Scatterer Detection}
In the case of an unqualified channel, the beam should be switched to the better direction, i.e., the target scatterer with maximum power should be detected. Hence, the issue can be considered a scatterer classification mission by classifying the scatterers into two classes, i.e., scatterer with maximum power $S_{max}$ (class: 1) and other scatterers (class: 0). The graph neural network (GNN) is constructed by utilizing the net-architecture in \cite{b14}. The GNN is the graph learning method for the graph data process. In which the $\mathbf {Aggregate}(\cdot)$ and $\mathbf {Combine}(\cdot)$ are the critical operators for modeling, where the former serves as the aggregation function of the neighborhood information, and the latter passes the aggregated node feature to a learnable layer to generate node embedding for the GNN layer.

\par Let $a_v^{(p)}$ stand for the nodes representing the structural information captured within the $p$-hop network neighborhood in $k$ iterations of aggregation. Hence, the $p$-th layer can be denoted by
\begin{equation}
a_v^{(p)}= \mathbf {Aggregate}^{(p)}(\{h_u^{(p-1)}:u \in \mathcal N(v)\}). 
\label{eq5}
\end{equation}
\begin{equation}
h_v^{(p)}= \mathbf {Combine}^{(p)}(h_v^{(p-1)},a_v^{(p)}),
\label{eq6}
\end{equation}
where $h_v^{(p)}$ is the feature vector of node $v$ at the $p$-th iteration, for $p=1,2,\cdots, P$ and $\mathcal N(v)$ is a set of nodes adjacent to $v$. The $h_v^{(0)}$ is initialized with $X_v$. As for the model, 8 MLPs are constructed. For each MLP, 6 hidden layers are deployed, where 1024 neurons are set for each layer. Finally, scores of two classes can be obtained by a fully-connected network.

However, the classification is independently deployed for each internal scatterer. For the unique detected scatterer of one environment, the scatterer classification results of a specific environment are ranked by the classification probability. In practice, the scatterer with the top class 1 probability is selected, or the scatterer with the minimum class 0 probability when all classified 0 is considered the final prediction result.

\section{Simulation and Results}

\subsection{Simulation Settings}
\par The environment samples with random scatterer changing are considered. The 3D modeling software Blender and the ray-tracing tool WirelessInSite are utilized for the traceable environment, and channel generation \cite{b10} as shown in Fig. 4. The propagation area's length, width, and height are set at 15, 10, and 3 m, and the Tx and Rx are set on the two sides of the diagonal. Regular scatterers with random numbers, positions, and dimensions are generated. The dataset includes samples with $J\in [3,12]$ numbers of internal scatterers.

\begin{figure}[!htb]
\centering
\includegraphics[width=0.38\textwidth]{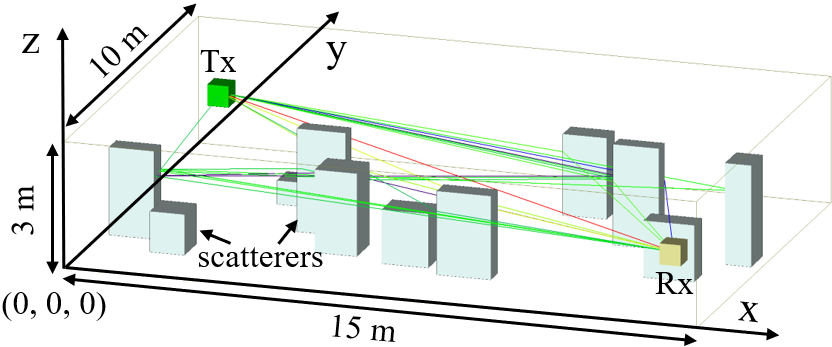}
\caption{A simulated sample with a random scatterer layout.}
\end{figure}
\par The training and testing data are a random selection of samples with different numbers of scatterers for generalization verification of the prediction method, in which the samples of the testing data consist of 4, 8, and 12 scatterers, and the rest samples are training data. The corresponding channels at 28 GHz are produced using the omnidirectional antenna, and six-order reflections and one-order diffraction are set. After selecting the NLOS samples, there are 1475 and 265 samples in training and testing data.

\subsection{Performance Metrics}
As for the binary classification problem, the precision and receiver operating characteristic curve (ROC) are given for performance analysis with the device of one NVIDIA GeForce RTX 2080. The precision $pre$ can be calculated as
\begin{equation}
pre = \rm{\frac{TP}{TP+TN}}
\label{eqa8}
\end{equation}
where true-positive (TP) and true-negative (TN) are the true samples classified as positive and negative. 

The ROC is plotted with a false-positive rate (FPR) and true-positive rate (TPR), which can be expressed as
\begin{equation}
\rm FPR = \rm{\frac{FP}{FP+TN}, TPR = \frac{TP}{TP+FN}},
\label{eqa8}
\end{equation}
where the false-positive (FP) and false-negative (FN) denote the false samples that be predicted with positive and negative. The crucial feature of ROC is the area under curve (AUC), where the closer the AUC is to 1, the better the performance.
\subsection{Quality and Target Scatterer Classification Results}
The quality threshold is set of 60\% cumulative distribution function (CDF) of received power. The SVM models with linear, polynomial, radial basis, and sigmoid kernel functions get 0.88, 0.92, 0.89, and 0.53 precisions, respectively. In which the polynomial kernel offers the best result. Then, according to the precision calculation, we can obtain the scatterer classification's accuracy is 0.89. The ROC and AUC of the SVM and GNN-based model are shown in Fig. 5, indicating the identity ability.

\begin{figure}[htbp]
    \centering
    \includegraphics[width=0.43\textwidth]{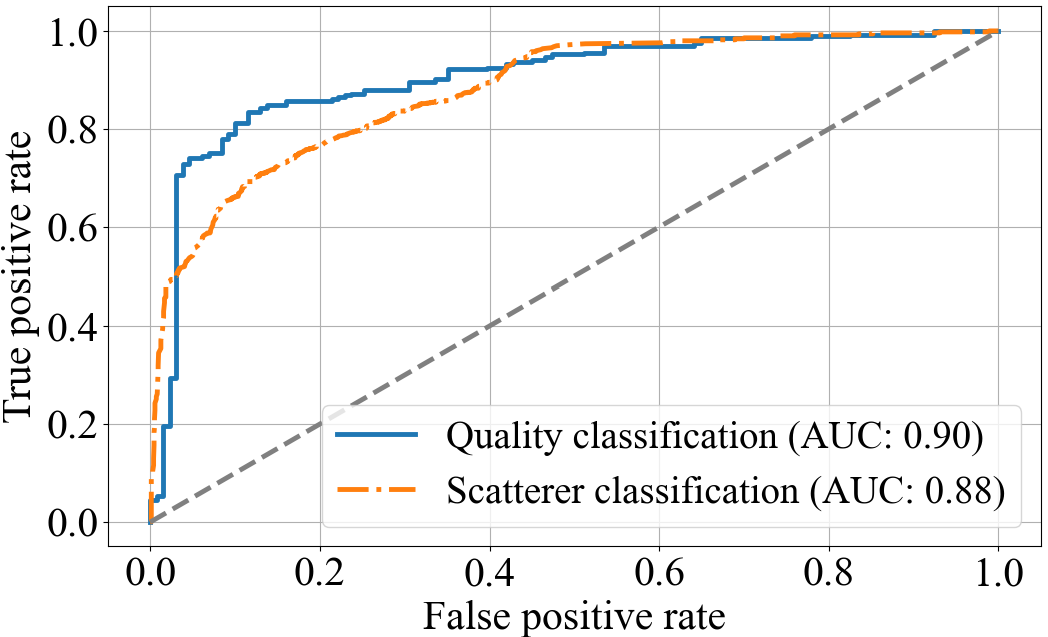}
    \caption{The ROC curve and AUC of quality and scatterer classification.}
	  \label{fig6}
\end{figure}

Moreover, based on the scatterer classification results, the target scatterer can be selected for an environment sample by the rank of classification scores. Moreover, the beam indices prediction in \cite{b4} is tested by utilizing the code at \cite{b15} for comparison. In which the top view images are generated by converting the coordinates of the dataset and setting the scatterers with diverse grayscale according to the different heights. The digital architecture system is employed with 8 antennas corresponding to 8 classes. The results are indicated in TABLE I. 

\begin{table}[h]
\caption{The Precision Comparisions}
\begin{center}
\begin{tabular}{|c|c|c|}
\hline
\textbf {Method} & \textbf {Configuration}  & \textbf{Precision} \\
\hline
Proposed Target Scatterer Detection & \textbackslash & 0.90 \\
\hline
\multirow{3}*{Beam Indices Prediction in [4]} & Top-1 & 0.51\\
\cline{2-3}
				  & Top-2 & 0.72\\
\cline{2-3}
				  & Top-3 & 0.84\\
\hline
\end{tabular}
\label{tab1}
\end{center}
\end{table}

The precision of the proposed method is around 0.9, while the top-3 precision of the method in \cite{b4} is around 0.84, which can hardly adapt to changing environments using few training data. Moreover, the testing time is compared for cost evaluation in TABLE II. The results illustrate that the proposed PESaBP method can save over 87\% time cost, which can support the online prediction for changing environments. 

\begin{table}[h]
\caption{The Comparisions of Testing Time}
\begin{center}
\begin{tabular}{|c|c|}
\hline
\textbf {Action} & \textbf {Testing time (ms)} \\
\hline
Proposed Channel Quality Evaluation & 4.7 \\
\hline
Proposed Target Scatterer Detection & 0.33 \\
\hline
Beam Indices Prediction in [4] & 41 \\
\hline
\end{tabular}
\label{tab1}
\end{center}
\end{table}

\section{Conclusion and Future Work}
This paper is interested in the PES definition, in which The PES is considered the task-oriented environment representation set according to the concerned channel properties. Therefore, the semantic mapping between the propagation environment and applications can be built directly. Simulation results of the PESaBP method indicate the efficiency and precision in NLOS scenarios, which have the potential to support online prediction in the ever-changing environment.


\begin{thebibliography}{1}
\bibitem{b1} P. Zhang, \emph{et al.}, “Ubiquitous-X: Constructing the future 6G networks,” \emph{SCIENTIA SINICA Informationis}, vol. 50, no. 6, pp 913-930, 2020.
\bibitem{b2} G. Nie, \emph{et al.}, “A predictive 6G network with environment sensing enhancement: From radio wave propagation perspective,” \emph{China Commun.}, vol. 19, no. 6, pp. 105-122, 2022.
\bibitem{b3} W. Xu, F. Gao, J. Zhang, X. Tao and A. Alkhateeb, “Deep learning based channel covariance matrix estimation with user location and scene images," \emph{IEEE Trans. on Commun.}, vol. 69, no. 12, pp. 8145-8158, 2021.
\bibitem{b4} M. Alrabeiah, A. Hredzak and A. Alkhateeb, “Millimeter wave base stations with cameras: vision-aided beam and blockage prediction," \emph{2020 91th VTC}, pp. 1-5, 2020.
\bibitem{b5} P. Zhang, \emph{et al.}, “Toward wisdom-evolutionary and primitive-concise 6G: A new paradigm of semantic communication networks," \emph{Engineering}, no. 8, pp. 60-73, 2022.
\bibitem{b6} H. Xie, \emph{et al.}, “Deep learning based semantic communications: An initial investigation," \emph{2020 21th GLOBECOM}, pp. 1-6, 2020.
\bibitem{b7} K. Niu, \emph{et al.}, “Towards semantic communications: A paradigm shift," arXiv preprint arXiv:2203.06692, 2022.
\bibitem{b8} J. Zhang, “The interdisciplinary research of big data and wireless channel: A cluster-nuclei based channel model,” \emph{China Commun.}, vol. 13, no. 2, pp 14-26, 2016.
\bibitem{b9}L. Yu, Y. Zhang, J. Zhang and Z. Yuan, “Implementation framework and validation of cluster-nuclei based channel model using environmental mapping for 6G communication systems,” \emph{China Commun.}, vol. 19, no. 4, pp 1-13, 2022.
\bibitem{b10} Y. Sun, \emph{et al.}, “Environment features-based model for path loss prediction,” \emph{IEEE Wireless Commun. Lett.}, vol. 11, no. 9, pp. 2010-2014, 2022.
\bibitem{b11} Z. Yuan, J. Zhang, Y. Zhang, P. Tang and L. Tian, “A Novel Complex PCA-based Wireless MIMO Channel Modeling Methodology,” \emph{2020 92th VTC}, pp. 1-5, 2020.
\bibitem{b12} Y. Sun, \emph{et al.}, “Environment information-based channel prediction method assisted by graph neural network”, \emph{China Commun.}, Accepted, 2022.
\bibitem{b13} T. M. Hoang, N. M. Nguyen and T. Q. Duong, “Detection of eavesdropping attack in UAV-aided wireless systems: Unsupervised learning with one-class SVM and k-means clustering," \emph{IEEE Wireless Commun. Lett.}, vol. 9, no. 2, pp. 139-142, 2020, 
\bibitem{b14} K. Xu, \emph{et al.}, “How powerful are graph neural networks?” arXiv preprint arXiv:1810.00826, 2018.
\bibitem{b15} M. Alrabeiah. [Online]. Available: https://github.com/malrabeiah/\par CameraPredictsBeams.
\end{thebibliography}
\end{document}